\documentclass[fleqn,usenatbib]{mnras}

\usepackage{newtxtext,newtxmath}
\usepackage{longtable} 
\usepackage{threeparttablex} 
\usepackage{subfigure}
\usepackage{multirow}
\usepackage[T1]{fontenc}

\DeclareRobustCommand{\VAN}[3]{#2}
\let\VANthebibliography\thebibliography
\def\thebibliography{\DeclareRobustCommand{\VAN}[3]{##3}\VANthebibliography}

\usepackage{graphicx}	% Including figure files
\usepackage{amsmath}	% Advanced maths commands
\usepackage{bm}

\usepackage{amssymb}	% Extra maths symbols

\title[LSBG Auto Detect]{Automatic detection of low surface brightness galaxies from SDSS images}

\author[Zhenping Yi et al.]{
	Zhenping Yi,$^{1}$\thanks{E-mail:yizhenping@sdu.edu.cn}
	Jia Li,$^{1}$
	Wei Du,$^{2}$\thanks{E-mail:wdu@nao.cas.cn}
	Meng Liu,$^{1}$
	Zengxu Liang,$^{1}$
	Yongguang Xing,$^{1}$
	Jingchang Pan,$^{1}$
	\newauthor
	Yude Bu,$^{1}$
	Xiaoming Kong$^{1}$ and
	Hong Wu$^{2}$\\
	$^{1}$School of Mechanical, Electrical and Information Engineering, Shandong University, 180 Wenhua Xilu, Weihai, 264200, China\\
	$^{2}$Key Laboratory of Optical Astronomy, National Astronomical Observatories, Chinese Academy of Sciences, 20A Datun Road, Chaoyang District, 100101, China\\
}

\date{Accepted XXX. Received YYY; in original form ZZZ}

\pubyear{2022}

\begin{document}
	\label{firstpage}
	\pagerange{\pageref{firstpage}--\pageref{lastpage}}
	\maketitle
	% Abstract of the paper
	\begin{abstract}

	Low surface brightness (LSB) galaxies are galaxies with central surface brightness fainter than the night sky. Due to the faint nature of LSB galaxies and the comparable sky background, it is difficult to search LSB galaxies automatically and efficiently from large sky survey. In this study, we established the Low Surface Brightness Galaxies Auto Detect model (LSBG-AD), which is a data-driven model for end-to-end detection of LSB galaxies from Sloan Digital Sky Survey (SDSS) images. Object detection techniques based on deep learning are applied to the SDSS field images to identify LSB galaxies and estimate their coordinates at the same time. Applying LSBG-AD to 1120 SDSS images, we detected 1197 LSB galaxy candidates, of which 1081 samples are already known and 116 samples are newly found candidates. The B-band central surface brightness of the candidates searched by the model ranges from 22 mag arcsec $^ {- 2} $ to 24 mag arcsec $^ {- 2} $, quite consistent with the surface brightness distribution of the standard sample. 96.46\% of LSB galaxy candidates have an axis ratio ($b/a$) greater than 0.3, and 92.04\% of them have $fracDev\_r$\textless 0.4, which is also consistent with the standard sample. The results show that the LSBG-AD model learns the features of LSB galaxies of the training samples well, and can be used to search LSB galaxies without using photometric parameters. Next, this method will be used to develop efficient algorithms to detect LSB galaxies from massive images of the next generation observatories.
	
	\end{abstract}
	
	% Select between one and six entries from the list of approved keywords.
	% Don't make up new ones.
	\begin{keywords}
		Low surface brightness galaxies, galaxies: fundamental parameters, methods: data analysis
	\end{keywords}
	
	%%%%%%%%%%%%%%%%%%%%%%%%%%%%%%%%%%%%%%%%%%%%%%%%%%
	
	%%%%%%%%%%%%%%%%% BODY OF PAPER %%%%%%%%%%%%%%%%%%
	
\section{Introduction}
 Low surface brightness (LSB) galaxies are conventionally defined as galaxies with central surface brightnesses fainter than the night sky (\cite{BothunImpey-109}). In most cases, the central region of a galaxy is the brightest part of the whole galaxy. Thus, a galaxy with its B-band central surface brightness $\mu_0(B)$ less than a certain threshold value is traditionally regarded as an LSB galaxy. Generally, the threshold value of $\mu_0(B)$ to classify galaxies as LSB galaxies in the literature varies from $\mu_0(B)\geq$23.0 mag arcsec$^{-2}$(\cite{BothunImpey-109}) to $\mu_0(B)\geq$22.0 mag arcsec$^{-2}$(\cite{BurkholderImpey-141}).
 
Compared with high surface brightness (HSB) galaxy, low surface brightness galaxies are generally rich in gas, low in metal abundance($\leq$ 1/3 solar abundance; \cite{McgaughSchombert-113}), dust-poor \cite[e.g.][]{MatthewsDriel-149}, and also have diffuse, low-density stellar disks \cite[e.g.][]{deBlokMcgaugh-150, BurkholderImpey-141, O'NeilBothun-125, TrachternachBomans-120}, indicating that LSB galaxies have low level of star formation activities and stay unevolved, especially in the disk. \cite[e.g.][]{DasReynolds-151, GalazHerrera-Camus-152}.
	
LSB galaxies are one of the main component of the realm of galaxies. In terms of number density, LSB galaxy may account for a large proportion in the universe (30\%\textasciitilde60\%, \cite[e.g.][]{McgaughSchombert-113, Mcgaugh-114, BothunImpey-109, ONeilBothun-117, TrachternachBomans-120, HaberzettlBomans-121, MartinKaviraj-137}). Therefore, the study of LSB galaxies is of great significance for our exploration of the universe. For example, the study of the spatial distribution of LSB galaxies helps to test the predictions of dark matter cosmology for large-scale structure forms \cite[e.g.][]{CutriSkrutskie-110}, and explore the formation process under low gas surface density environment (\cite{BothunImpey-111}), as well as the density of baryonic matter and the formation and evolution of galaxies (\cite{BraineHerpin-112}), etc.

However, the contribution of LSB galaxies to the universe has long been underestimated because they are dim compared to the night sky and hard to spot (\cite{BothunImpey-109,CeccarelliHerrera-Camus-122}). This problem has been concerned, and many related studies have been done. For example, Impey and his colleagues scanned the UK Schmidt plates using the Automated Plate Measuring(APM) mechanism and found 693 LSB galaxies, which formed the most extensive catalog of LSB galaxies at that time (\cite{ImpeySprayberry-107}). Their team then studied the properties of these LSB galaxies in a series of works. \cite{O'NeilBothun-106} first discovered the red LSB cluster in Texas Survey. \cite{MonnierRagaigneVanDriel-96, MonnierRagaignevanDriel-59, MonnierRagaignevanDriel-60} selected a sample of about 3,800 LSB galaxies from the all-sky near-infrared (NIR) Two Micron all sky measurement (2MASS), then obtained their 21cm HI observations, and estimated Hi masses of LSB galaxies subsamples. \cite{KniazevGrebel-148}  developed a software to search for LSB galaxy from images of SDSS combined with SDSS photometry information, tested their method using sample data provided by \cite{ImpeySprayberry-107}, and finally obtained 129 LSB galaxies. From the galaxy samples of SDSS DR4 database, \cite{ZhongLiang-105} established a large sample of 12,282 forward low brightness galaxies, and \cite{Liang2010} statistically studied the spectral properties of these samples. \cite{DuWu-145} and \cite{he2020} detected 1129 face-on LSB galaxies and 281 edge-on LSB galaxies from $\alpha$.40 SDSS DR7  (\cite{Haynes-156}) and studied the LSB galaxy properties of their samples, respectively.

In most previous studies, the detection of LSB galaxies usually directly measures the brightness of the central plane of galaxies. The detection consists of the following steps: estimating the sky background of the image and subtracting it; selecting sources like galaxies from the images; measuring the surface brightness, and determining whether the source is an LSB galaxy. however, due to the low surface brightness of LSB galaxies, the sky background subtraction bias may lead to the loss of LSB galaxies, especially for fainter ones. For example,  \cite{DuWu-145}, \cite{Adelman_McCarthy_2008}, and \cite{LiuFS2008} pointed out that SDSS overestimates the skylight to a certain extent, which leads to the underestimation of galaxy flux and has a great influence on the photometric results of fainter galaxies. Besides, it is difficult to perform automatic and an accurate measurement, because some processing is more complex and may need a manual operation (\cite{he2020}). Therefore, it is very necessary to study a new automatic search method for LSB galaxies.

In recent years, machine learning, especially artificial neural network, has been more and more applied to astronomical images to cope with the increasing astronomical data both in size and complexity. For example, \cite{10.1093/mnras/stz2816}, \cite{10.1093/mnras/staa501}, \cite{10.1093/mnras/stab1552} developed machine learning models for automated morphological classification of galaxies.  \cite{2019AAS...23314430B} trained a Convolutional Neural Network model to predict the gas-phase metallicity of galaxies using $gri$ images from SDSS. \cite{Pasquet:2018qlv} developed a Deep 
Convolutional Neural Network to estimate photometric redshifts for galaxies in SDSS. Deep learning techniques are also applied to find galaxy–galaxy strong gravitational lens (\cite{10.1093/mnras/stx1665}).
In the above researches, classification methods or regression methods were used to derive categories or attributes of celestial objects, in which an image contains one major object and corresponds to a label or an attribute value that is used to represent the content of an image. In the field of astronomy, small images containing celestial objects cut from a whole astronomical image are often used for classification or regression tasks. Compared with classification or regression tasks, object detection is more complex but more powerful, which is one of the computer vision tasks. In an object detection task, the categories and positions of multiple objects in an image are recognized at the same time, including both classification and regression sub tasks. Traditional object detection approaches include scanning the whole image using a sliding window, extracting features by using techniques like SIFT ( \cite{1999Object}), HOG (\cite{2005Histograms}), then using algorithms, such as SVM (\cite{1995Support}), for the classification of target objects. Neural network approaches of object detection is able to tackle some drawbacks of traditional approaches, because the deep learning architectures are able to learn more complex features. Besides, object detection using neural network techniques are able to do end-to-end object detection without specifically defining features. Therefore, object detection based on deep learning performs outstanding in many applications of images processing.

In this paper, we present a low surface brightness galaxy auto-detect (LSBG-AD) model based on deep learning, to search for LSB galaxies from SDSS field images. By learning the features of LSB galaxies in training samples, LSBG-AD locates and identifies LSB galaxies directly from SDSS images. The photometry image contains the most abundant information of LSB galaxies, such as extended shape, disk-like or irregular shape, which are very important for the detection of low brightness galaxies. Thus, searching for LSB galaxies from images can maximize the use of these features. Moreover, LSB galaxies can be recognized and located directly from astronomical images using this method which is an end-to-end information extraction process. It is very convenient to search LSB galaxies from a large number of photometry images without extracting photometric parameters from the images first.

This paper outlines the methods used in searching LSB galaxies from SDSS images. In Section 2, we introduce LSB galaxy samples used in this work and data preprocessing. Our LSBG-AD model is described in detail in Section 3. The performance and the searching results are described in Section 4. In Section 5, we give an analysis of candidates and other test results, as well as the caveats of the model. Lastly, in Section 6 we provide a summary of our work.
	
    \section{DATA}
	\subsection{Data Sample}

	The data we used as standard sample to build the LSBG-AD model include 1129 LSB galaxies from \cite{DuWu-145}(hereafter Du2015 sample). These LSB galaxies were selected from the $\alpha$.40 SDSS DR7 sample. ALFALFA (the Arecibo Legacy Fast ALFA survey; \cite{2005AJ....130.2598G}; \cite{2007NCimB.122.1097G}; \cite{2007NCimB.122.1109H}) is a very wide area blind extragalactic Hi survey. The Sloan Digital Sky Survey or SDSS (\cite{1998AJ....116.3040G}; \cite{YorkAdelman-147}; \cite{2001ASPC..238..269L}) is a wide-field optical/infrared imaging and spectroscopy survey using a dedicated 2.5-m wide-angle optical telescope at Apache Point Observatory. SDSS DR7 (\cite{2009ApJS..182..543A}) is the seventh major data release and provides images, spectra, and scientific catalogs and shares some footprints with the $\alpha$.40 data set, which covers about 40\% of the full ALFALFA survey sky area(\cite{Haynes-156}).
	Using $\alpha$.40-SDSS DR7 as parent sample, Du et al. defined a sample of 1129 LSB galaxies with ${{\mu }_{0}}$(B) > 22.5 mag arcsec$^{-2}$ and the axis ratio $b/a >$0.3. This sample is a relatively unbiased sample of gas-rich and disk-dominated LSB galaxies, spanning from 22.5 mag arcsec$^{-2}$ to 28.2 mag arcsec$^{-2}$ in $\mu_0(B)$.
	
	\begin{figure}
		\centering
		\subfigure[]{
			\begin{minipage}[r-band]{0.4\textwidth}
			\includegraphics[width=1\textwidth]{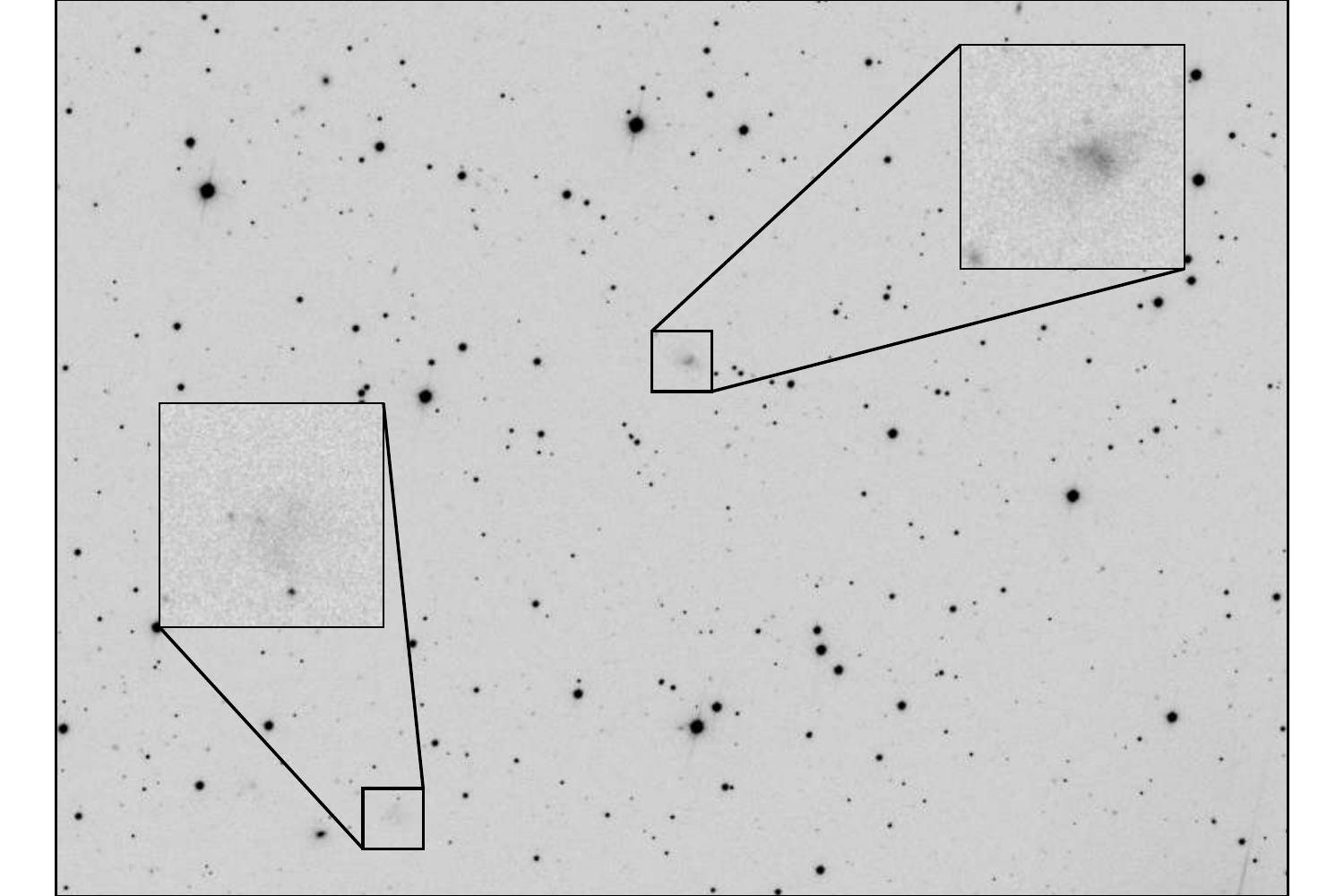}
			\end{minipage}
		}
		\subfigure[]{
			\begin{minipage}[g-band]{0.4\textwidth}
			\includegraphics[width=1\textwidth]{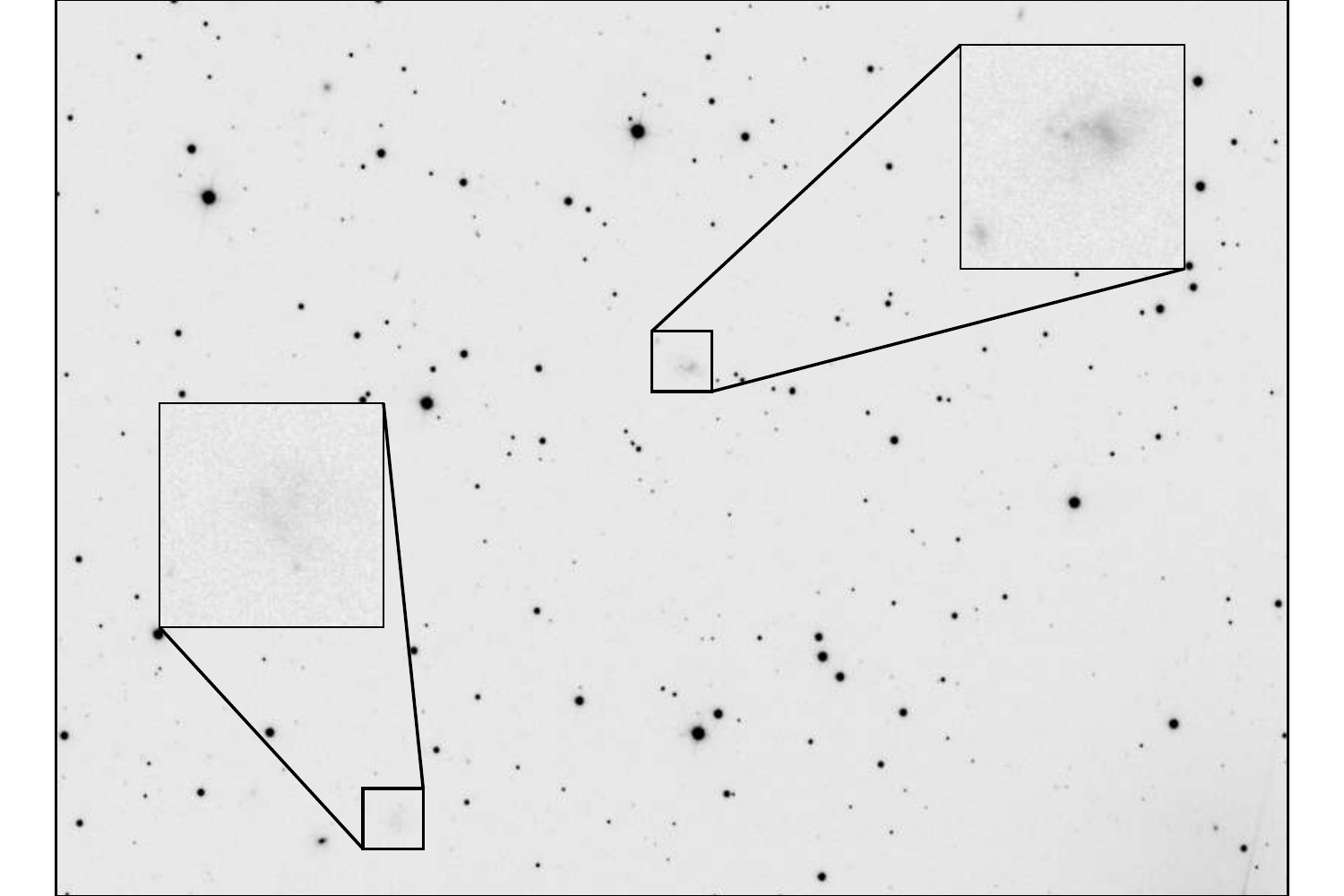}
			\end{minipage}
		}
		\caption{Two LSB galaxies in an SDSS field. (a) and (b) is g and r band images, The LSB galaxies are marked with small boxes, also enlarged and displayed in the big boxes}\label{fig:LSBGbox1}
	\end{figure}
	
	Since we aim to search for LSB galaxies from SDSS images, our training data also includes the SDSS images containing 1129 LSB galaxies. According to R.A. and Dec. coordinates of 1129 LSB galaxies, 1120 (Some images contain two LSB galaxies) SDSS images have been selected from the SDSS DR7 image database. These downloaded images are corrected frames (fpC-images) that have been preprocessed by the SDSS photometric pipeline, including bias corrections, flat field, corrections for bad pixels by interpolated values and sky subtraction. The size of a corrected SDSS image is 2048 * 1489 pixels, covering an approximately 9.83'*13.52' sky area. 
	SDSS Images were taken using a photometric system of five filters (u, g, r, i, and z). Because the g-band and r-band SDSS images contain the most abundant information of LSB galaxies, they are often used to search for LSB galaxies in previous studies. Therefore, we also choose these two band images as training data.

	Then the images were randomly divided into a training set containing 1000 images and a test set containing 120 images. An example of g-band and r-band SDSS images with two LSB galaxies marked with boxes are shown in Figure \ref{fig:LSBGbox1}.
	
		\begin{figure}
		% \centering
		\includegraphics[width=8.5cm]{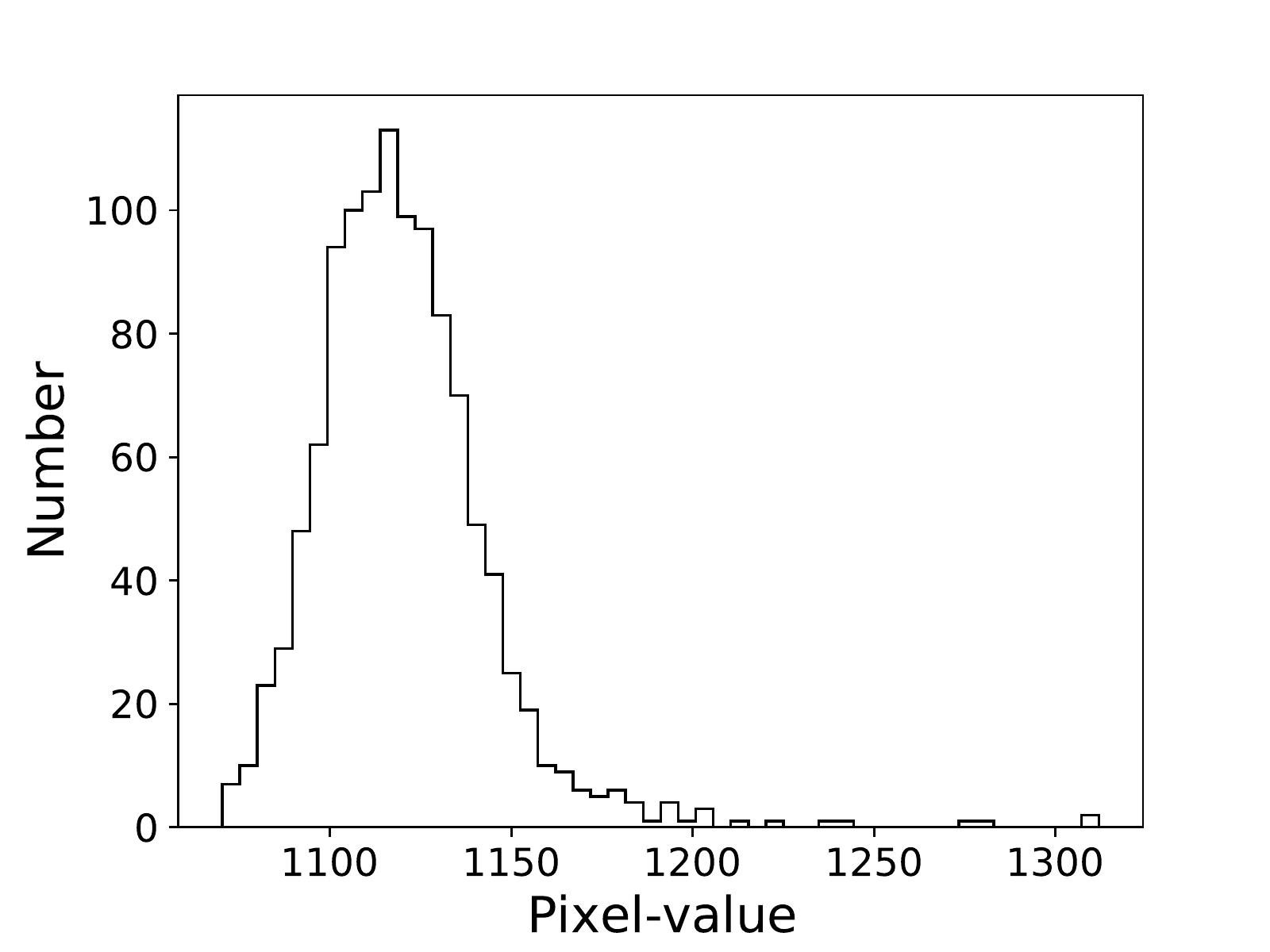}
		\caption{The distribution of the mean values of the center 8*8 pixels of LSB galaxies. The values of most LSB galaxy pixels are in the range of 1000-1200.}
		\label{fig:pixeldis}
	\end{figure}

	\begin{figure*}
		\includegraphics[width=11cm]{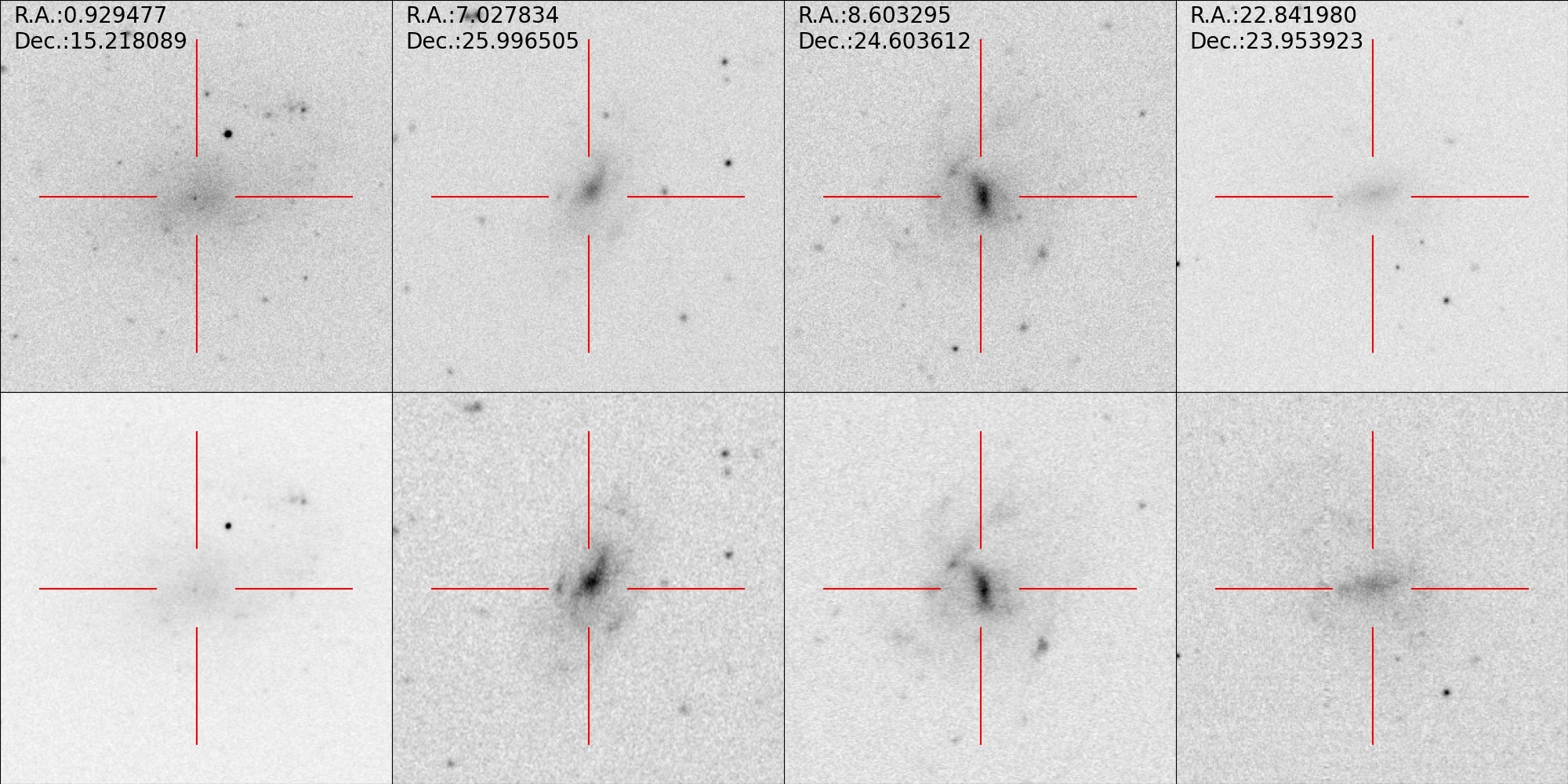}
		\caption{The four LSB galaxies in the training data, each column of images represents an LSBG, r-band image is on the top, and g-band image is at the bottom (the image has been processed by the pre-processing method in 2.2)}\label{fig:example1}
	\end{figure*}

	\subsection{Data preprocessing}	
	There are some offsets between a g-band image and its corresponding r-band image in SDSS data, we aligned them using Reproject [https://reproject.readthedocs.io/en/stable/] by referring to r frame coordinates.
	We investigated the distribution of pixel values in SDSS images and found that the pixel values span  in the range of 0-65,535. Due to the low brightness of LSB galaxies, their pixel values are low, most in the range of 1000-1200, as shown in Figure \ref{fig:pixeldis}, which shows the distribution of the mean values of the center 8*8 pixels of 1129 LSB galaxies. To remove redundancy, let the neural network pay more attention to LSB galaxies, we cut the range of the data using formula \ref{qua:formula1}.  We set the values below 1000 to 1000, and the values above 1255 to 1255 \footnote{The trials with clipping thresholds of 1200, 1255, 1500 and 2000 were conducted and 1255 worked best in our case.}. Then we subtracted 1000 from all the values, and finally, the range of pixels is 0-255. In Figure \ref{fig:example1}, the SDSS images of four LSB galaxies after numerical scale processing are shown. \\

	\begin{equation}
		f(x)=\left\{
		\begin{aligned}
			0      &  &x\le1000 \\
			x-1000 &  &1000<x \le 1255 \\
			255    &  &x>1255
		\end{aligned}
		\right\}\label{qua:formula1}
	\end{equation}
	
	\section{Building LSBG-AD model}
	LSBG-AD is a deep neural network model for fast detection of LSB galaxies from SDSS images. It identifies the presence and location of one or more LSB galaxies in a given SDSS image. It is a challenging problem that LSBG-AD involves building upon methods for distinguishing LSB galaxies from the other galaxies, stars, and backgrounds, and meanwhile predicting their location. The problem is a typical computer object detection task, which is not a single classification task or a single regression task, but a combination of the two. 
	
    LSBG-AD is a one-stage object detection model, similar to the popular YOLO (You only look once, \cite {RedmonDivvala-140}) model. In the field of computer vision, there are two kinds of deep learning object detection detectors: two-stage and one-stage detectors. Two-stage detectors carry out detection tasks in two steps: proposing a set of regions of interest first and then classifying the region candidates, such as R-CNN family (\cite {Girshick-74, Girshick2014, RenHe-65}). In contrast, a one-stage detector skips the region proposal stage and runs detection directly over a dense sampling of possible locations, such as the typical YOLO series (\cite {RedmonDivvala-140, RedmonFarhadi-78, RedmonFarhadi-79}). LSBG-AD model is similar to that of YOLO in theory, but due to the characteristics of astronomical image data and LSB galaxies, special architecture and settings are adopted.
    In the LSBG-AD model, the input image can be considered to be divided into 24 x 32 grids of cells. For each grid, the prediction of whether there is an LSB galaxy in the grid will be made, including the class probability p and the coordinates (x,y) of the LSB galaxy.
    
    The establishment of LSBG-AD is divided into three steps: firstly, a deep neural network is constructed; Secondly, the loss function is defined to evaluate the quality of the model; Finally, through the learning process, we search and find the best parameters in the network parameter space and optimize the performance of the model. The three steps are described in detail below.
    
    \subsection{Network architecture}		
\begin{figure*}
		\centering
		\includegraphics[height=18.0cm, width=15cm]{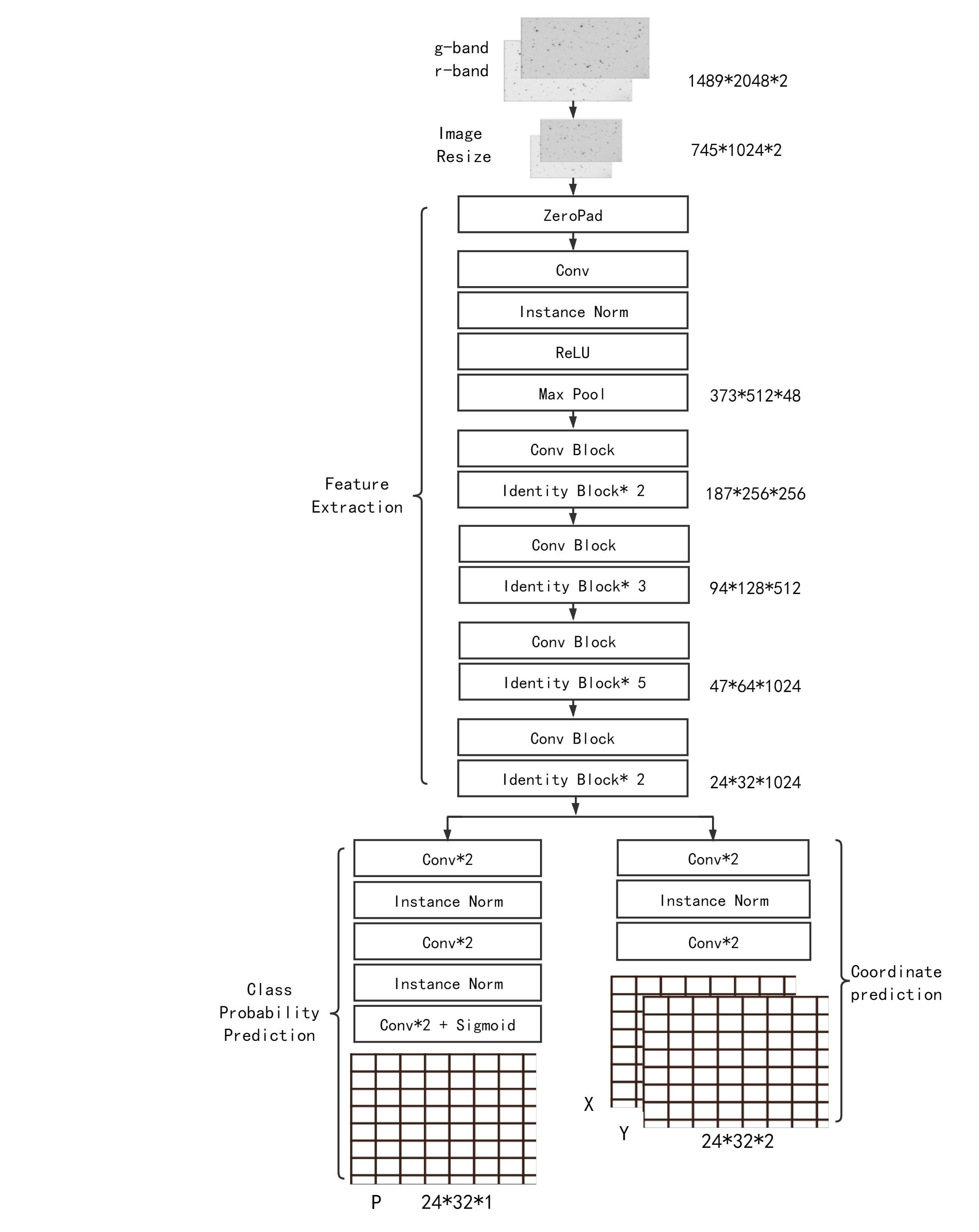}
		\caption{Network architecture of the LSBG-AD. The architecture Of LSBG-AD consists of four components:image input module, feature extraction module, class probability prediction module, and coordinate prediction module. LSBG-AD takes the g-band and r-band images of SDSS as the input and outputs the coordinate position and class probability of the LSB galaxies detected from the images. } \label{fig:Structure}
	\end{figure*}

	The network architecture Of LSBG-AD consists of four modules: image input module, feature extraction module, class probability prediction module, and coordinate prediction module. The network architecture is shown in Figure \ref{fig:Structure}. 
	
	\begin{itemize} 
		\item[1.] Image input module. This module accepts g-band and r-band images that have been processed following section 2.2.  Subsequently, the images can be adjusted to the appropriate size according to the computing performance and capacity of the computer. In this work,  1489 pixels * 2048 pixels matrix corresponding to the g-band and r-band images, are resized to the half size of 745 pixels * 1024 pixels. 
		
		\item [2.] Feature extraction module. The design of this module refers to the feature extraction structure of a residual neural network ResNet50 (\cite{he2016identity}). ResNet50 is a kind of convolutional neural network (CNN, \cite{hinton2012improving, o2015introduction} ) of 50 layers deep. Different from the ordinary convolutional neural network, residual neural networks utilize skip connections, or shortcuts to jump over some layers. The advantage of adding skip connection is to avoid the problem of vanishing gradients or mitigate the degradation problem. In other words, residual neural networks ease the training of deeper networks. Here, we use a feature extraction structure that is almost the same as the first 49 layers of ResNet50. The difference is that fewer filters were used while doing convolution operations because of the performance limitation of our computer. In general, the module adopts 49 layers feature extraction structure composed of convolutional layer (Conv), Conv Block (each includes 3 layers), Identity Block(each includes 3 layers). This feature extraction structure finally transforms 745 * 1024 * 2 image to 24 * 32 * 1024 feature map. 
		
		\item [3.] Class probability prediction module. This module uses a convolutional neural network consisting of 6 convolutional layers to integrate the features of the feature map coming from the previous module and fit these features to Class probability. Finally, the module outputs the 24 * 32 class probabilities matrix, in which $p_{ij}$ represents the probability of an LSB galaxy locating in the grid cell $(i,j)$ of the original image.
	
		\item [4.] Coordinate prediction module. This module also uses convolutional neural networks to predict the center coordinates matrix X, Y of LSB galaxies, ($x_{ij}, y_{ij}$) represents the center coordinates of an LSB galaxy detected from grid cell $(i,j)$.
	\end{itemize}
    
    \subsection{Loss Function of LSBG-AD}
	In deep learning, a loss function is often defined to reflect the difference or error between the prediction and the actual value. Therefore, it can evaluate how well the neural network models the data set. The value of a loss function is also called loss. In the optimization process, when the loss function reaches the minimum, the network best models the data set and gets the optimal performance.
	
	The loss function of LSBG-AD consists of two components, One component describes the loss of class probability p and the other describes the loss of coordinate position. The loss function is defined as formula 2:
	
	\begin{equation}
		\begin{split}
			Loss = &-\Sigma_i^{24}\Sigma_j^{32}(p_{ij}\log{\hat{p}_{ij}}+(1-p_{ij})\log{(1-\hat{p}_{ij}}) \\
			&+\Sigma_i^{24}\Sigma_j^{32}\hat{p}_{ij}((x_{ij}-\hat{x}_{ij})^2+(y_{ij}-\hat{y}_{ij})^2)^{1/2}
		\end{split}
	\end{equation}
	
	where $\hat{p}_{ij}$ and $p_{ij}$ respectively represent the predicted class probability and the real class probability of an LSB galaxy in a cell$(i,j)$. ($\hat{x}_{ij}$,$\hat{y}_{ij}$) and ($x_{ij}$,$y_{ij}$) respectively represent the predicted center coordinates and the real center coordinates of an LSB galaxy in cell$(i,j)$. The first component describing the loss of class probability uses the cross-entropy loss which is often used in the binary class problem, and the second component describing the loss of coordinate position uses a modified square loss by multiplying a factor of  $\hat{p}_{ij}$ \footnote{A hyper parameter $\lambda$ is usually multiplied to the loss of coordinates. Here it is set to 1, which is determined by hyper parameter optimization trails.}. The modified squared loss measures the distance between the predicted position and the real position, meanwhile reducing the function of coordinates of objects with low probability in the cell.
	
	\subsection{Learning Process}
	Finally, the data was fed to the network, and through iterative learning, the network gradually learned the ability to identify LSB galaxies. We used stochastic gradient descent to learn the optimal parameters of the network with a gradient clipping setting for suppressing the gradient explosion in the learning process. After 100 epochs of training, the model converges and we get the basic model that can identify LSB galaxies.
	
	\subsection{Searching Strategy}
    When we use the basic model for large-scale search, we may encounter a problem. That is, the center coordinate of some LSB galaxies may fall on the boundary of the cell, which may lead to detection failure or inaccurate location of the target. In order to solve this problem, we use a strategy, that is shifting the image and putting it into the network together with the original image, and finally, remove duplicates and keep the most reasonable LSB galaxies from multiple results of one SDSS image. The specific steps are described as follows.
	\begin{itemize}
		\item[1.] First, the original SDSS image was sent to the basic network and a batch of results was obtained.
		\item[2.] Shift each image by cutting out 16 columns of pixels (a quarter width of a cell) on the right side of the image and adding 16 columns of 0 values on the left side of the image. Then send the shifted images into the basic network to obtain the second batch of results.
		\item[3.] Shift each image by cutting the first 16 rows of pixels (about a quarter height of a cell)  on the top of the image and adding 0-pixel value to the bottom 16 rows. Then send the shifted images into the basic network to obtain the third batch of results.
	    \item [4.] Shift each image by cutting the top 16 rows of pixels and the left 16 columns of pixels, adding 0-pixel value to the bottom 16 rows and the right 16 columns, that is doing the step 2 and step 3 simultaneously. Then send the shifted images into the basic network to obtain the fourth batch of results.
		\item[5.] Remove duplicate sources from four batches of results.
		If the distance between two sources detected in the same image is greater than a threshold distance, they are considered independent sources. If the distance between two sources is less than the threshold, they may be the same source. The source with higher class probability is kept, while the sources with lower class probabilities are discarded. Here the threshold distance is set to 10 arc seconds.
	\end{itemize}

	\section{Results}
	
	\subsection{Performance Evaluation}
	As an important output of the LSBG-AD model, p represents the class probability of an LSB galaxy in a cell of an image. For each SDSS image, there are 24 * 32 p values corresponding to 24 * 32 cells. According to the class probability  p, we selected the candidates of LSB galaxies.
	
	A class probability threshold, usually 0.5, can be chosen to distinguish LSB candidates. Here, we chose the threshold of 0.9 to obtain LSB galaxies with a higher confidence level. Those with a class probability greater than 0.9 were identified as candidates of LSB galaxies.
	
	We then used a measure $recall$ to evaluate the performance of the LSBG-AD model. After training the model using the training dataset and predicting the class probability of LSB galaxies in the testing data, some LSB galaxies have been obtained. Among detected LSB galaxies, the already known ones in the training data set or test data set are called true positives ($TP$). $Recall$ is defined as
	\begin{equation}
	 recall=\frac{TP}{N}	
	\end{equation}
	where $N$ is the total number of positives that are already known LSB galaxies in the training set or testing set.

	\subsection{Searching results}
	We used the LSBG-AD model and the searching strategy described in Section 3.4 to search all the images, including the training set and the test set.
	
\begin{figure}  
		\centering
		\includegraphics[width=8.5cm,height=6cm]{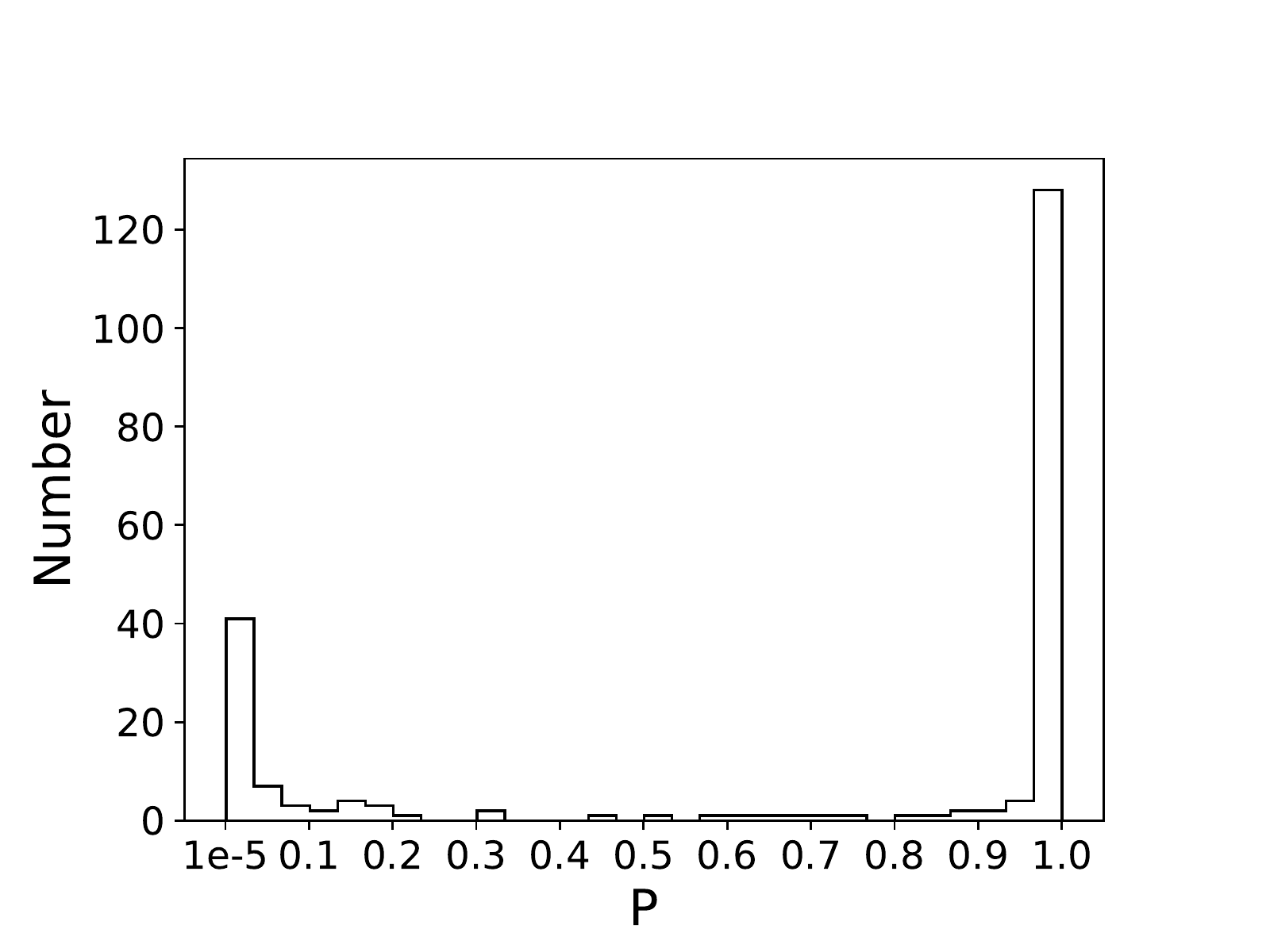}
		\caption{Class probability p distribution of testing data. A class probability close to 1 means that the detected source is an LSB galaxy with high confidence. Note that the class probability  p $< 10^{-5}$ is not included in the distribution. These small values are numerous, because most class probability p of cells in images are close to 0, indicating that  there are no LSB galaxies in the grid, but only the background or other objects.}\label{fig:Pvalue}
	\end{figure}

	From 120 test images, a total of 120*24*32 class probability p were obtained. Most p values are close to zero ($< 10^{-5}$), which indicates there are probably no LSB galaxies in their corresponding cells. Thus, we did not involve these low p values in Figure \ref{fig:Pvalue}, in which the remaining class probability ($>=10^{-5}$) distribution of 120 test images were shown. The class probabilities of 134 sources are greater than 0.9, indicating these sources likely to be LSB galaxies with high confidence. There is a small group of probability values that are lower than 0.1 and the remaining few fall between 0.1 and 0.9. This distribution indicates that for most of the sources, it is definite to determine whether it is an LSB galaxy.
	
\begin{figure*}
		\centering
		\includegraphics[width=13cm]{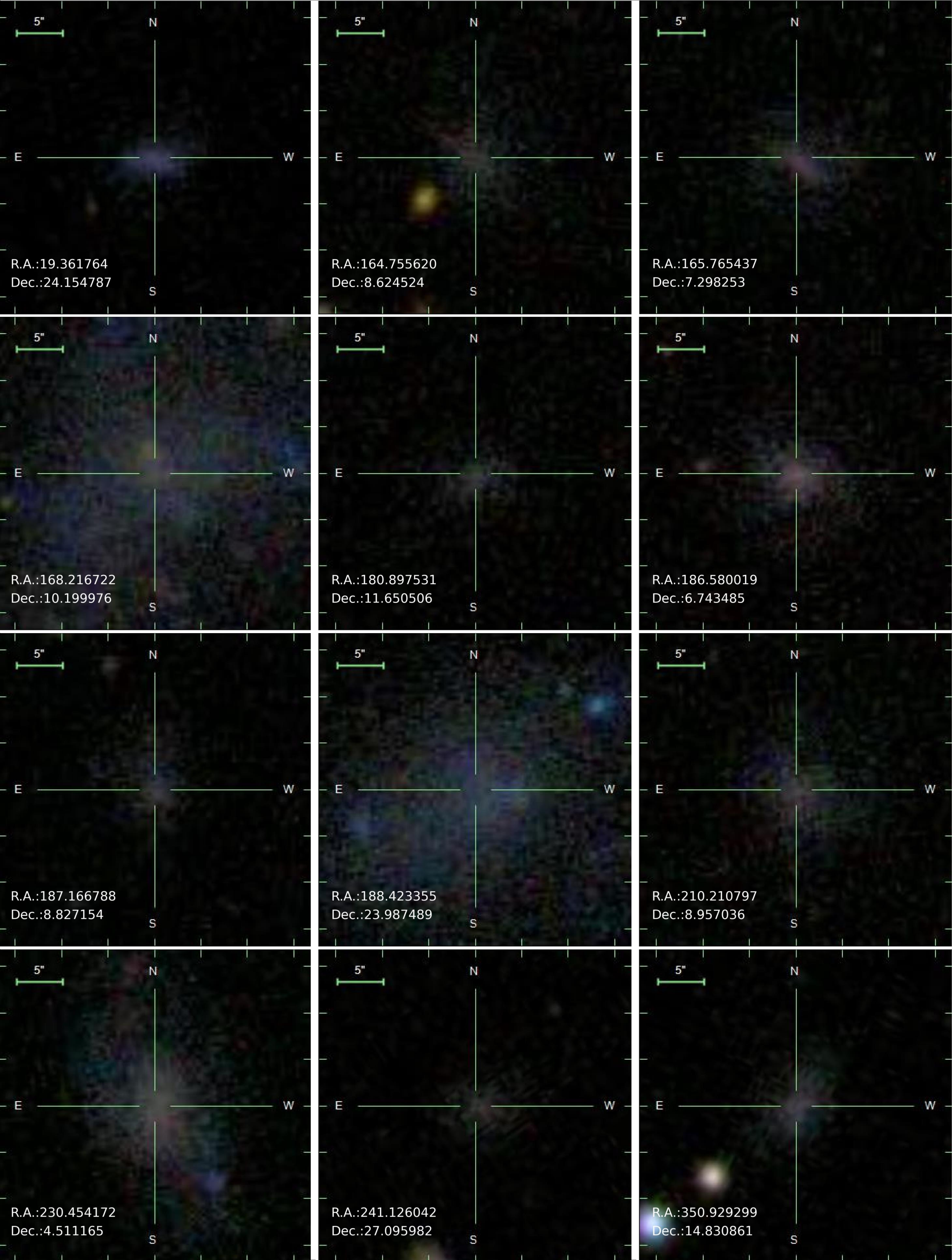}
		\caption{LSB galaxy candidates detected by LSBG-AD model. The coordinate R.A. and Dec. are the center of candidates predicted by LSBG-AD}\label{fig:candidate}
	\end{figure*}

 From 120 test images, 134 LSB galaxy candidates were found, and among them, 112 ones were already labeled in the testing set and 22 ones were newly found candidates. From 1,000 training images, 1,063 LSB galaxies candidates were found, of which 969 ones were already known in the training set and 94 ones were newly found by the model, as shown in Table \ref{Tab:results}. For the training set, the recall is 96.23\%, and for the test set, the recall is 91.8\%. The recall of the testing set reflects the performance of our model applied to new SDSS images. In Figure \ref{fig:candidate} we show 12 newly detected LSB galaxy candidates from testing images.
 
Our experiments are performed using Python 3.6 on an NVIDIA GTX 1660 GPU. Using our searching strategy, detecting one SDSS image spent 3 seconds on the platform.
	
	\begin{table*}
		\centering
		\caption{Summary of searching results using LSBG-AD.}
		\label{Tab:results}
		\begin{tabular}{ccccccc}\hline
			&Image &Labeled &Model found &Overlapped &Model newly-found &Recall\\\hline
			Training set &1000  &1007    &1063  &969        &94        &96.23\%\\
			Test set     &120   &122     &134   &112        &22        &91.8\%\\
			Total        &1120  &1129    &1197  &1081       &116       &95.75\%\\\hline
		\end{tabular}
	\end{table*}

	\section{DISCUSSION}
	\subsection{Analysis of candidates}
	
    For LSB galaxy candidates, we want to obtain their B-band central surface brightness for evaluating our detection. However, it is difficult to obtain accurate central surface brightness. It is related to the parameters such as magnitude, disk scale length, axis ratio and redshift. The uncertainty of these parameters will affect the accuracy of central surface brightness of LSB galaxy. We obtained the photometric parameters of these LSB galaxy candidates from SDSS catalog to calculate their central surface brightness. However, due to the uncertainty of the parameters, the central surface brightness calculated by using SDSS photometric parameters may not be accurate, which is used here to reflect the surface brightness distribution of the whole sample.
    
    We calculate the B-band central surface brightness according to the following formula:
    
	\begin{equation}
		\begin{split}
			\mu_0 = m + 2.5\log_{10}(2\pi\alpha^2q) -10\log_{10}(1+z)\\
		\end{split}
	\end{equation}  
    
    \begin{equation}
		\begin{split}
			\mu_0(B)=\mu_0(g)+0.47( \mu_0(g)-\mu_0(r)) + 0.17 
		\end{split}
	\end{equation}

	where m is magnitude measured by SDSS photometry ($expMag\_r$, $expMag\_g$) and subsequently corrected by galactic extinction ($extinction\_r$, $extinction\_g$), $\alpha$ is the disk-sacle length ($rs$), converted from disk half-light radius ($Re$, SDSS $expRad\_r$, $expRad\_g$) ( $rs = Re/1.678$).  $q$ is the axis ratio ($expAB\_r$, $expAB\_g$), and $z$ is the redshift ( SDSS $z$ ).
	
\begin{figure*}
		\centering
		\includegraphics[width=12cm]{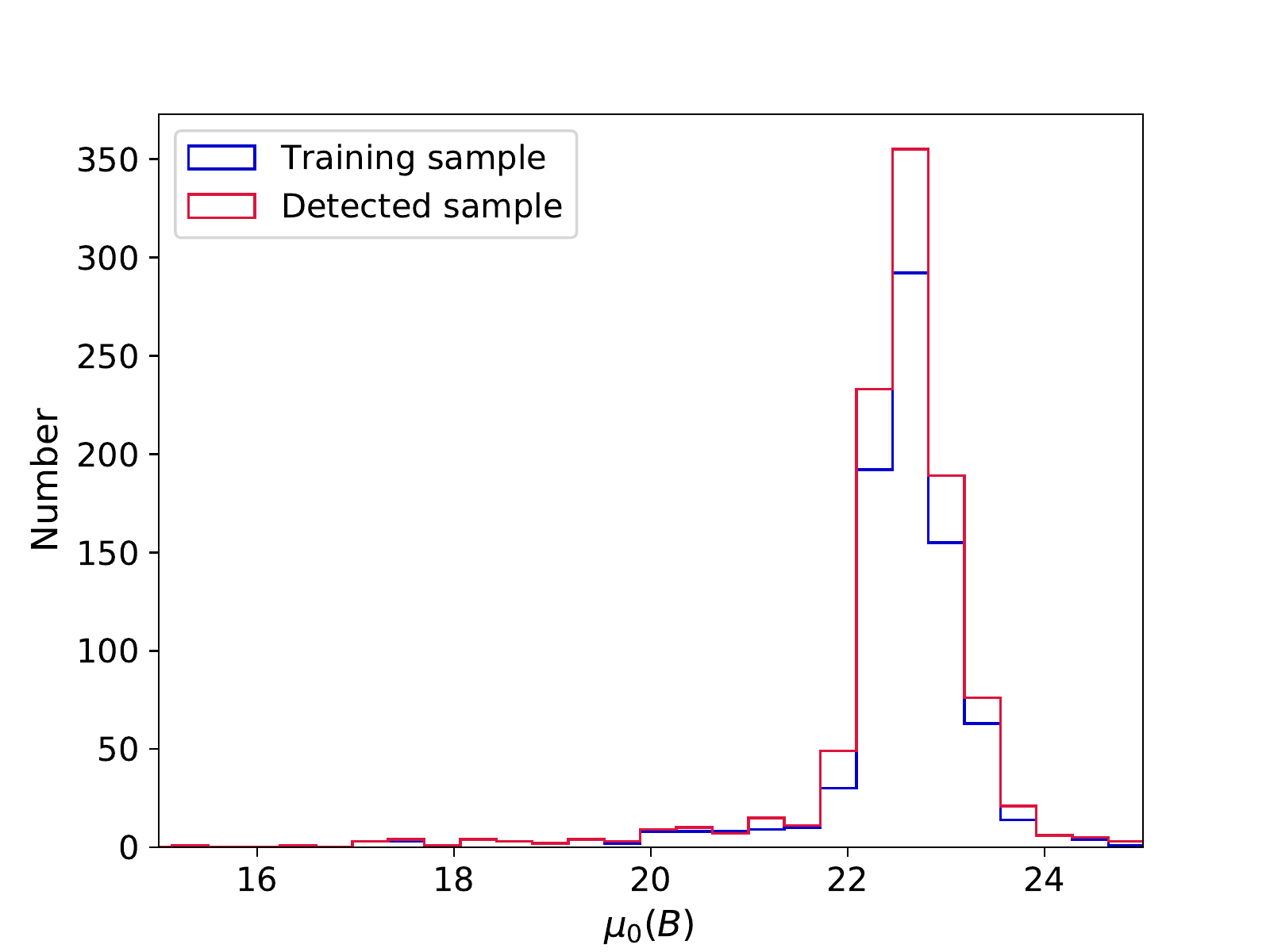}
		\caption{Distribution of B-band central surface brightness of LSB galaxy candidates(red) and training samples(blue).\label{fig:Ubrightness}}
	\end{figure*}

	We cross-matched these candidates with SDSS photometric catalog within a 5 arcsec radius to obtain photometric parameters from SDSS catalog. First, we removed the sources with abnormal $expRad$ value (about 8\% values are too small) through visual inspection. Then we used formula (4) to calculate $\mu_0(g)$ and $\mu_0(r)$ respectively, and used formula (5) to calculate the B-band central surface brightness. Finally, except for a few LSB candidates (absent from the SDSS catalog or missing the red shift or with abnormal $expRad$ value), we measure the B-band central surface brightness for rest candidates. The central surface brightness $\mu_0(B)$ distribution is shown in red in Figure \ref{fig:Ubrightness}. For comparison, we also calculated the B-band central surface brightness of LSB galaxies in Du2015 sample according to SDSS photometry, and their distribution is shown in blue in Figure \ref{fig:Ubrightness}. The surface brightness of the LSB candidates is mainly distributed between 22-24 mag arcsec$^{-2}$, which is the same as that of Du2015 sample. It should be pointed out that $\mu_0(B)$ of the Du2015 sample is in the coverage of 22.5 mag arcsec$^{-2}$ to 28.3 mag arcsec$^{-2}$, however, there is a deviation between the surface brightness calculated by using SDSS photometry and the measurement of Du2015. $\mu_0(B)$ using SDSS photometry has the deviation of  about 0.5 mag arcsec$^{-2}$ lower than that of Du2015's. Therefore, in this work of using the photometirc parameters of SDSS, it is more appropriate to choose galaxies darker than 22 mag arcsec$^{-2}$ as the LSB galaxies , which is also a frequently adopted standard( e.g. \cite{ZhongLiang-105}, \cite{Boissier2003}).

	In order to learn about the morphological properties of these newly detected LSB candidates, We used the parameter $fracDeV\_r$ selected from SDSS catalog to quantify the galaxy profile type ($fracDeV\_r$ = 1 for a de Vaucouleurs profile; $fracDeV\_r$ = 0 for an exponential profile). It shows that 92.04\% of the candidates have $fracDeV\_r$<$0.4$, indicating that most of the candidates were disk-dominated galaxies. Then axis ratio $b/a$ of the candidates was calculated, and it was found that 96.46\% of the candidates have $b/a>0.3$, belonging to face-on galaxies. For comparison, 94.2\% of the training samples are with $fracDeV\_r<0.4$, and 100\% of the training samples are with $b/a>0.3$. The results show that the candidates are consistent with the training samples in shape.
	
\begin{figure*}
		\centering
		\includegraphics[width=13.3cm]{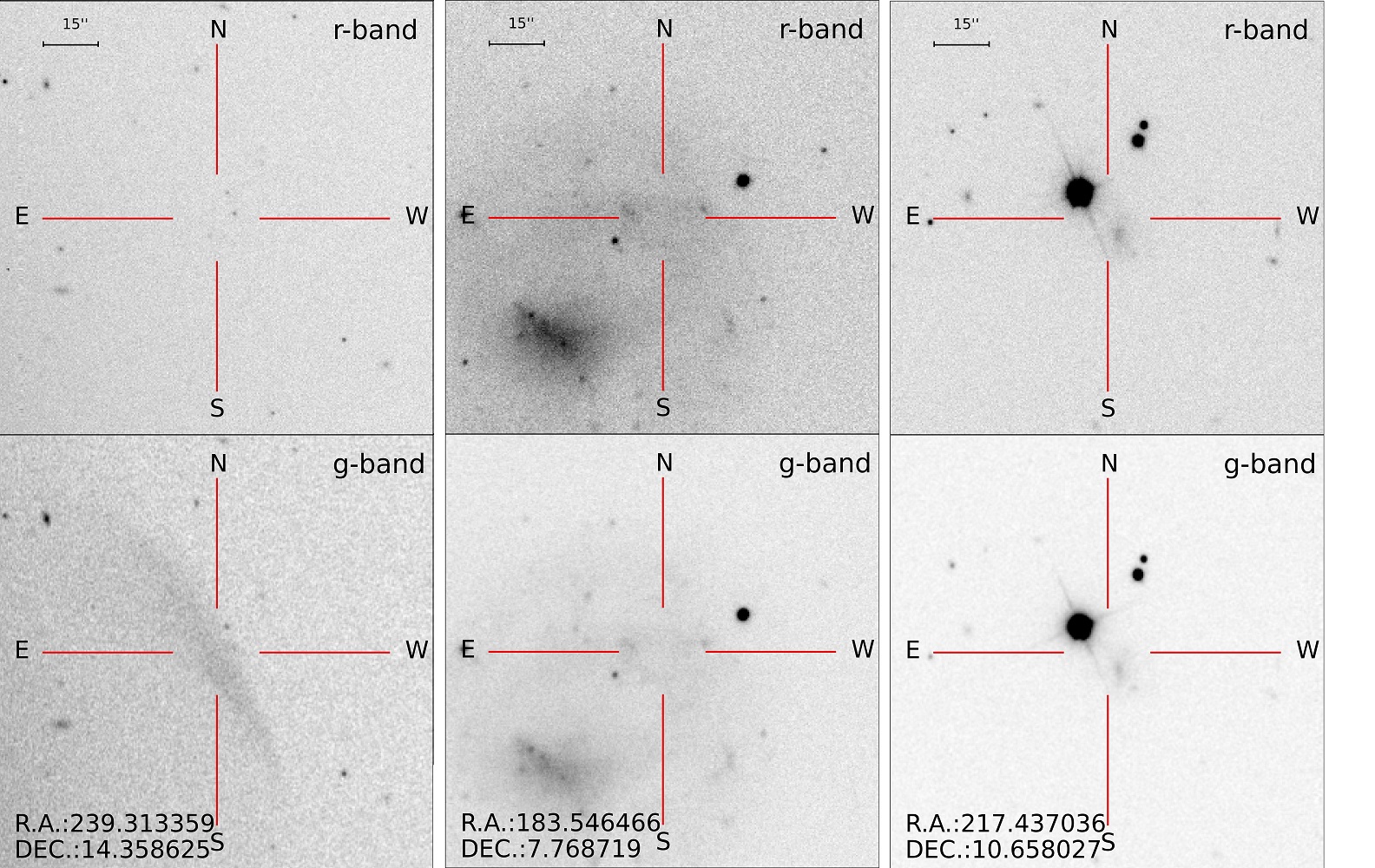}
		\caption{unmatched candidates with SDSS objects within a 5 arcsec radius. \label{fig:missed}}
	\end{figure*}

	There are 3 unmatched candidates with SDSS objects within a 5 arcsec radius. Figure \ref{fig:missed} presents the images of 3 LSB galaxy candidates respectively in r-band and g-band. In the first column, there is an extended shape that only appeared in g-band but not in r-band, it is an artifact that may be caused by the bright star nearby. In the second column, there is a larger deviation in the predicted coordinate, because the star-forming region or knot in the arm of a large spiral galaxy is regarded as the center of an LSB galaxy. In the third column, the detection is disturbed by a nearby bright star, thus the predicted coordinates of the LSB galaxy have a large deviation with its real center.
	
	\subsection{Test using another sample}
\begin{figure*}
		\centering
		\includegraphics[width=12cm]{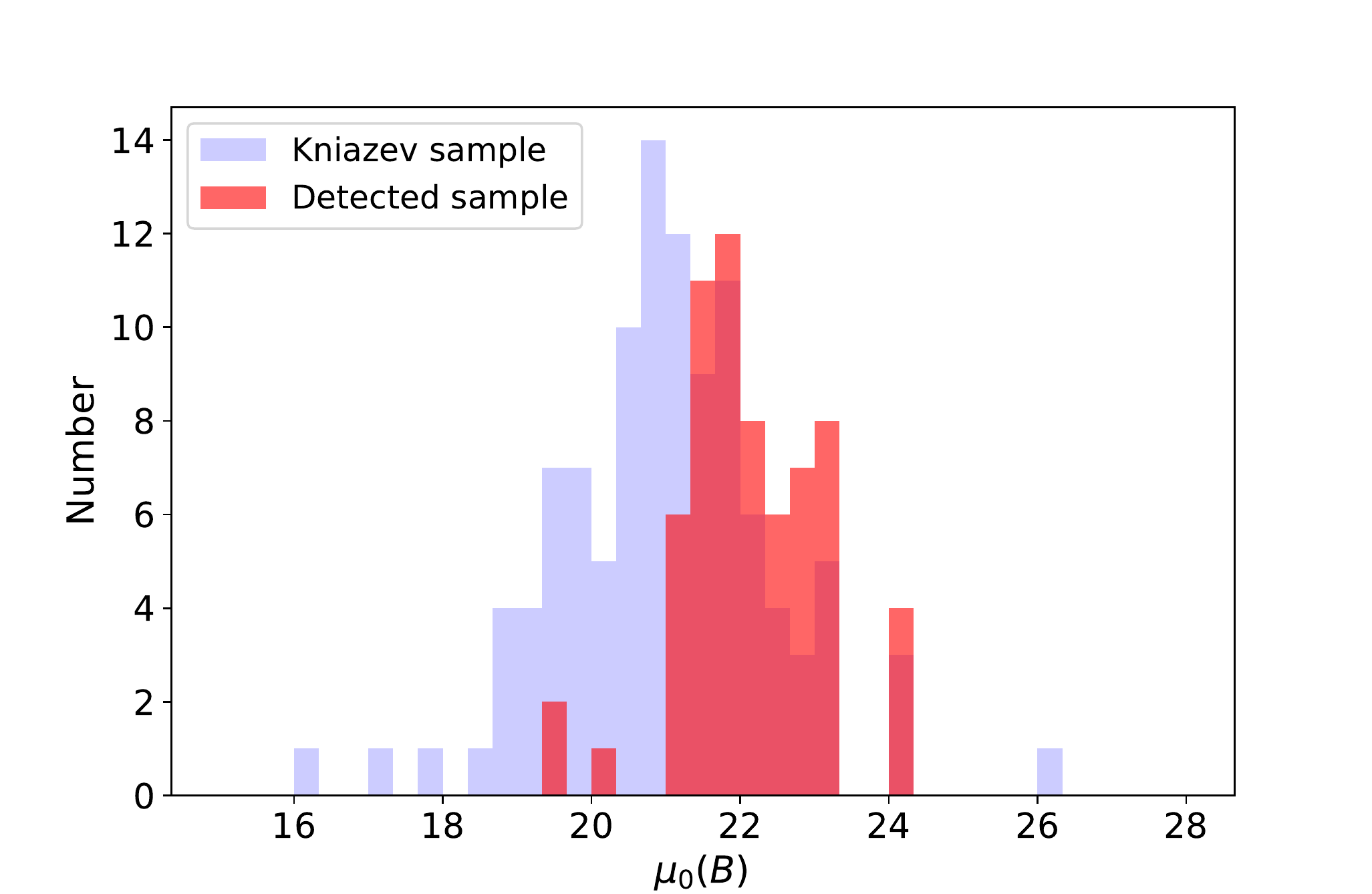}
		\caption{The distribution of B-band central surface brightness of detected LSB galaxy candidates (red) and Kniazev2004 sample (\protect\cite{KniazevGrebel-148}) (blue).}\label{fig:Kniazev}
	\end{figure*}

	We also used another sample to test our model. The LSBG-AD model was applied to search for LSB galaxies on an SDSS image sample collecting from \cite{KniazevGrebel-148} (hereafter Kniazev2004 sample). Kniazev et al. developed a new photometry software and detected 129 galaxies from 93 SDSS images. We applied the LSBG-AD model to this sample and from 93 SDSS images, we detected 78 LSB galaxy candidates, 49 of which are also in the Kniazev2004 sample. According to the photometric parameters of SDSS, we calculated the B-band central surface brightness of the kniazev2004 sample and the candidates we detected, as shown in Figure \ref{fig:Kniazev}. The $\mu_0(B)$ for Kniazev samples is distributed in 16.06-26.02 mag arcsec$^ {- 2} $ with the average 20.95 mag arcsec$^{- 2} $, while the candidates we detected are distributed in 19.46-24.24 mag arcsec$^{- 2} $, with the average 22.14 mag arcsec$^{- 2} $. The samples we searched are darker. In the kniazev2004 sample, there are 23 candidates which $\mu_0(B)$ are darker than 22 mag arcsec$^{- 2} $. Our model successfully searched 22 of them, with the recall rate of 96\%. In addition, we also found 10 additional candidates whose $\mu_0(B)$ of the B-band are darker than 22mag arcsec$^{- 2} $. The testing results showed that our LSBG-AD model has an excellent ability to search LSB galaxy darker than 22 mag arcsec$^{-2}$ of $\mu_0(B)$ from SDSS images.

	\subsection{Model Limitations}
	There are 2 limitations of the current LSBG-AD model.
	\begin{itemize}
		\item[1.] The model may miss a few LSB galaxies. For example, ten LSB galaxies were missed during searching test set. We checked their images and found these sources are either too small and faint, or there are bright sources around, causing failed detection. In essence, the features of these special samples are not well learned by the model because of the small number of these special samples in the training data set. 
		\item[2.] In addition, there are some sources of contamination in the candidates, as mentioned in Section 5.1. The common sources of contamination are nearby bright stars, artifacts, knots or star-forming areas in large spiral galaxies. Besides, for some detected LSB galaxies, there may be a little deviation in coordinate position due to the interference of nearby bright stars.
	\end{itemize}

	\section{Conclusion}
	In this paper, we aim to develop an effective and automatic method to search for low surface brightness (LSB) galaxies in SDSS images. By using the 1129 LSB galaxies of Du2015 sample, we have built LSBG-AD, an LSB galaxies detection model based on a deep neural network, to achieve our goal.  LSBG-AD was successfully applied to 1,120 SDSS images to search for LSB galaxies. 1081 Of 1129 known LSB galaxies were recovered, with the training recall of 96.23\% and the test recall of 91.8\%. Furthermore, we used Kniazev2004 sample to test the LSBG-AD model. For darker than 22 mag arcsec$^{-2}$ LSB galaxies, the recall is 96\% . The test results show LSBG-AD has an excellent ability to search LSB galaxies with $\mu_0(B)$ greater than 22 mag arcsec$^{-2}$  from SDSS images.
	
	LSBG-AD uses the deep learning object detection technology to learn features of low surface brightness galaxies from SDSS images, such as shape, brightness and color, so it can automatically detect LSB galaxies without obtaining photometric parameters. Detecting LSB galaxies from images is the most direct way, which avoids the complex processing process of traditional surface brightness measurement. It is very effective in the search of important celestial candidates. Like most other supervised learning methods, LSBG-AD depends heavily on training data. A reliable, complete and balanced training data can reduce the limitation of such algorithms, which is also the direction for us to improve LSBG-AD in the future. A continuation of this work is applying LSBG-AD to search for LSB galaxies from massive images and more deep images, so as to provide large and rich samples for the research of LSB galaxies. 
	
	\section{ACKNOWLEDGMENTS}
	This research was supported by the National Natural Science Foundation of China (Grant Nos. U1931109, U1931209, 12090041, 12090040, 11873037, 11733006, 11803016), Cultivation Project for LAMOST Scientific Payoff and Research Achievement of CAMS-CAS and the Youth Innovation Promotion Association, Chinese Academy of Sciences. This research was partly supported by the GHfund B (20210702)
	
	We would like to thank the SDSS team for the wonderful released SDSS images and parameter catalog. Funding for the SDSS has been provided by the Alfred P. Sloan Foundation, the Participating Institutions, the National Science Foundation, the US Department of Energy, the NASA, the Japanese Monbukagakusho, the Max Planck Society, and the Higher Education Funding Council for England. The SDSS Web Site is http://www.sdss.org. 
	
	Software:astropy (\cite{robitaille2013streicher,2018}), python3 (\cite{van2009python}), pandas (\cite{wes2010proc}), numpy (\cite{van2011numpy, virtanen2020scipy}), matplotlib(\cite{4160265}),  tensorflow (\cite{tensorflow2015-whitepaper}), opencv (\cite{howse2013opencv})).

	%%%%%%%%%%%%%%%%%%%%%%%%%%%%%%%%%%%%%%%%%%%%%%%%%%
	\section*{Data Availability}
	
	The data underlying this article is at https://github.com/worldoutside/LSBG.

	%%%%%%%%%%%%%%%%%%%% REFERENCES %%%%%%%%%%%%%%%%%%
	% The best way to enter references is to use BibTeX:ref
	
	\bibliographystyle{mnras}
	\bibliography{lsbg} % if your bibtex file is called example.bib

\begin{thebibliography}{}
\makeatletter
\relax
\def\mn@urlcharsother{\let\do\@makeother \do\$\do\&\do\#\do\^\do\_\do\%\do\~}
\def\mn@doi{\begingroup\mn@urlcharsother \@ifnextchar [ {\mn@doi@}
  {\mn@doi@[]}}
\def\mn@doi@[#1]#2{\def\@tempa{#1}\ifx\@tempa\@empty \href
  {http://dx.doi.org/#2} {doi:#2}\else \href {http://dx.doi.org/#2} {#1}\fi
  \endgroup}
\def\mn@eprint#1#2{\mn@eprint@#1:#2::\@nil}
\def\mn@eprint@arXiv#1{\href {http://arxiv.org/abs/#1} {{\tt arXiv:#1}}}
\def\mn@eprint@dblp#1{\href {http://dblp.uni-trier.de/rec/bibtex/#1.xml}
  {dblp:#1}}
\def\mn@eprint@#1:#2:#3:#4\@nil{\def\@tempa {#1}\def\@tempb {#2}\def\@tempc
  {#3}\ifx \@tempc \@empty \let \@tempc \@tempb \let \@tempb \@tempa \fi \ifx
  \@tempb \@empty \def\@tempb {arXiv}\fi \@ifundefined
  {mn@eprint@\@tempb}{\@tempb:\@tempc}{\expandafter \expandafter \csname
  mn@eprint@\@tempb\endcsname \expandafter{\@tempc}}}

\bibitem[\protect\citeauthoryear{Abadi et~al.,}{Abadi
  et~al.}{2015}]{tensorflow2015-whitepaper}
Abadi M.,  et~al., 2015, {TensorFlow}: Large-Scale Machine Learning on
  Heterogeneous Systems, \url {https://www.tensorflow.org/}

\bibitem[\protect\citeauthoryear{{Abazajian} et~al.,}{{Abazajian}
  et~al.}{2009}]{2009ApJS..182..543A}
{Abazajian} K.~N.,  et~al., 2009, \mn@doi [\apjs]
  {10.1088/0067-0049/182/2/543}, \href
  {https://ui.adsabs.harvard.edu/abs/2009ApJS..182..543A} {182, 543}

\bibitem[\protect\citeauthoryear{Adelman-McCarthy et~al.,}{Adelman-McCarthy
  et~al.}{2008}]{Adelman_McCarthy_2008}
Adelman-McCarthy J.~K.,  et~al., 2008, \mn@doi [The Astrophysical Journal
  Supplement Series] {10.1086/524984}, 175, 297

\bibitem[\protect\citeauthoryear{{Boada} \& {Wu}}{{Boada} \&
  {Wu}}{2019}]{2019AAS...23314430B}
{Boada} S.,  {Wu} J.,  2019, in American Astronomical Society Meeting Abstracts
  \#233. p. 144.30

\bibitem[\protect\citeauthoryear{Boissier, Monnier~Ragaigne, Prantzos, van
  Driel, Balkowski  \& O'Neil}{Boissier et~al.}{2003}]{Boissier2003}
Boissier S.,  Monnier~Ragaigne D.,  Prantzos N.,  van Driel W.,  Balkowski C.,
   O'Neil K.,  2003, \mn@doi [Monthly Notices of the Royal Astronomical
  Society] {10.1046/j.1365-8711.2003.06703.x}, 343, 653

\bibitem[\protect\citeauthoryear{Bothun, Impey, Malin  \& Mould}{Bothun
  et~al.}{1987}]{BothunImpey-111}
Bothun G.~D.,  Impey C.~D.,  Malin D.~F.,   Mould J.~R.,  1987, The
  Astronomical Journal, 94, 23

\bibitem[\protect\citeauthoryear{Bothun, Impey  \& McGaugh}{Bothun
  et~al.}{1997}]{BothunImpey-109}
Bothun G.,  Impey C.,   McGaugh S.,  1997, Publications of the Astronomical
  Society of the Pacific, 109, 745

\bibitem[\protect\citeauthoryear{Braine, Herpin  \& Radford}{Braine
  et~al.}{2000}]{BraineHerpin-112}
Braine J.,  Herpin F.,   Radford S.,  2000, Astronomy and Astrophysics, 358,
  494

\bibitem[\protect\citeauthoryear{Burkholder, Impey  \& Sprayberry}{Burkholder
  et~al.}{2001}]{BurkholderImpey-141}
Burkholder V.,  Impey C.,   Sprayberry D.,  2001, Astronomical Journal, 122,
  2318

\bibitem[\protect\citeauthoryear{Cavanagh, Bekki  \& Groves}{Cavanagh
  et~al.}{2021}]{10.1093/mnras/stab1552}
Cavanagh M.~K.,  Bekki K.,   Groves B.~A.,  2021, \mn@doi [Monthly Notices of
  the Royal Astronomical Society] {10.1093/mnras/stab1552}, 506, 659

\bibitem[\protect\citeauthoryear{Ceccarelli, Herrera-Camus, Lambas, Galaz  \&
  Padilla}{Ceccarelli et~al.}{2012}]{CeccarelliHerrera-Camus-122}
Ceccarelli L.,  Herrera-Camus R.,  Lambas D.~G.,  Galaz G.,   Padilla N.~D.,
  2012, Monthly Notices of the Royal Astronomical Society: Letters, 426, L6

\bibitem[\protect\citeauthoryear{Cheng et~al.,}{Cheng
  et~al.}{2020}]{10.1093/mnras/staa501}
Cheng T.-Y.,  et~al., 2020, \mn@doi [Monthly Notices of the Royal Astronomical
  Society] {10.1093/mnras/staa501}, 493, 4209

\bibitem[\protect\citeauthoryear{Cortes \& Vapnik}{Cortes \&
  Vapnik}{1995}]{1995Support}
Cortes C.,  Vapnik V.,  1995, Machine Learning, 20, 273

\bibitem[\protect\citeauthoryear{Cutri et~al.,}{Cutri
  et~al.}{2000}]{CutriSkrutskie-110}
Cutri R.~M.,  et~al., 2000, Explanatory Supplement to the 2MASS Second
  Incremental Data Release

\bibitem[\protect\citeauthoryear{Dalal \& Triggs}{Dalal \&
  Triggs}{2005}]{2005Histograms}
Dalal N.,  Triggs B.,  2005, in IEEE Computer Society Conference on Computer
  Vision \& Pattern Recognition.

\bibitem[\protect\citeauthoryear{Das, Reynolds, Vogel, Mcgaugh  \&
  Kantharia}{Das et~al.}{2009}]{DasReynolds-151}
Das M.,  Reynolds C.~S.,  Vogel S.~N.,  Mcgaugh S.~S.,   Kantharia A.,  2009,
  The Astrophysical Journal

\bibitem[\protect\citeauthoryear{Du, Wu, Lam, Zhu, Lei  \& Zhou}{Du
  et~al.}{2015}]{DuWu-145}
Du W.,  Wu H.,  Lam M.~I.,  Zhu Y.,  Lei F.,   Zhou Z.,  2015, The Astronomical
  Journal, 149, 199

\bibitem[\protect\citeauthoryear{Galaz, Herrera-Camus, Lambas  \&
  Padilla}{Galaz et~al.}{2011}]{GalazHerrera-Camus-152}
Galaz G.,  Herrera-Camus R.,  Lambas D.~G.,   Padilla N.,  2011, Trustees of
  the British Museum,

\bibitem[\protect\citeauthoryear{{Giovanelli}}{{Giovanelli}}{2007}]{2007NCimB.122.1097G}
{Giovanelli} R.,  2007, \mn@doi [Nuovo Cimento B Serie]
  {10.1393/ncb/i2008-10442-9}, \href
  {https://ui.adsabs.harvard.edu/abs/2007NCimB.122.1097G} {122, 1097}

\bibitem[\protect\citeauthoryear{{Giovanelli} et~al.,}{{Giovanelli}
  et~al.}{2005}]{2005AJ....130.2598G}
{Giovanelli} R.,  et~al., 2005, \mn@doi [\aj] {10.1086/497431}, \href
  {https://ui.adsabs.harvard.edu/abs/2005AJ....130.2598G} {130, 2598}

\bibitem[\protect\citeauthoryear{Girshick}{Girshick}{2015}]{Girshick-74}
Girshick R.,  2015, Fast r-cnn

\bibitem[\protect\citeauthoryear{Girshick, Donahue, Darrell  \& Malik}{Girshick
  et~al.}{2014}]{Girshick2014}
Girshick R.,  Donahue J.,  Darrell T.,   Malik J.,  2014, in 2014 IEEE
  Conference on Computer Vision and Pattern Recognition. pp 580--587,
  \mn@doi{10.1109/CVPR.2014.81}

\bibitem[\protect\citeauthoryear{{Gunn} et~al.,}{{Gunn}
  et~al.}{1998}]{1998AJ....116.3040G}
{Gunn} J.~E.,  et~al., 1998, \mn@doi [\aj] {10.1086/300645}, \href
  {https://ui.adsabs.harvard.edu/abs/1998AJ....116.3040G} {116, 3040}

\bibitem[\protect\citeauthoryear{Haberzettl, Bomans  \& Dettmar}{Haberzettl
  et~al.}{2007}]{HaberzettlBomans-121}
Haberzettl L.,  Bomans D.~J.,   Dettmar R.-J.,  2007, Astronomy \&
  Astrophysics, 471, 787

\bibitem[\protect\citeauthoryear{{Haynes}}{{Haynes}}{2007}]{2007NCimB.122.1109H}
{Haynes} M.~P.,  2007, \mn@doi [Nuovo Cimento B Serie]
  {10.1393/ncb/i2008-10447-4}, \href
  {https://ui.adsabs.harvard.edu/abs/2007NCimB.122.1109H} {122, 1109}

\bibitem[\protect\citeauthoryear{Haynes}{Haynes}{2011}]{Haynes-156}
Haynes M. P. G.~R.,  2011, Astronomical Journal

\bibitem[\protect\citeauthoryear{He, Zhang, Ren  \& Sun}{He
  et~al.}{2016}]{he2016identity}
He K.,  Zhang X.,  Ren S.,   Sun J.,  2016, in European conference on computer
  vision. pp 630--645

\bibitem[\protect\citeauthoryear{He, Wu, Du, Liu, Lei, Zhao  \& Zhang}{He
  et~al.}{2020}]{he2020}
He M.,  Wu H.,  Du W.,  Liu H.-y.,  Lei F.-j.,  Zhao P.-s.,   Zhang B.-q.,
  2020, \mn@doi [The Astrophysical Journal Supplement Series]
  {10.3847/1538-4365/ab8ead}, 248, 33

\bibitem[\protect\citeauthoryear{Hinton, Srivastava, Krizhevsky, Sutskever  \&
  Salakhutdinov}{Hinton et~al.}{2012}]{hinton2012improving}
Hinton G.~E.,  Srivastava N.,  Krizhevsky A.,  Sutskever I.,   Salakhutdinov
  R.~R.,  2012, arXiv preprint arXiv:1207.0580

\bibitem[\protect\citeauthoryear{Howse}{Howse}{2013}]{howse2013opencv}
Howse J.,  2013, OpenCV computer vision with python.
Packt Publishing Ltd

\bibitem[\protect\citeauthoryear{Hunter}{Hunter}{2007}]{4160265}
Hunter J.~D.,  2007, \mn@doi [Computing in Science Engineering]
  {10.1109/MCSE.2007.55}, 9, 90

\bibitem[\protect\citeauthoryear{Impey, Sprayberry, Irwin  \& Bothun}{Impey
  et~al.}{1996}]{ImpeySprayberry-107}
Impey C.~D.,  Sprayberry D.,  Irwin M.~J.,   Bothun G.~D.,  1996, The
  Astrophysical Journal Supplement Series, 105, 209

\bibitem[\protect\citeauthoryear{Kniazev, Grebel, Pustilnik, Pramskij,
  Kniazeva, Prada  \& Harbeck}{Kniazev et~al.}{2004}]{KniazevGrebel-148}
Kniazev A.~Y.,  Grebel E.~K.,  Pustilnik S.~A.,  Pramskij A.~G.,  Kniazeva
  T.~F.,  Prada F.,   Harbeck D.,  2004, Astronomical Journal, 127, 704

\bibitem[\protect\citeauthoryear{Lanusse, Ma, Li, Collett, Li, Ravanbakhsh,
  Mandelbaum  \& Póczos}{Lanusse et~al.}{2017}]{10.1093/mnras/stx1665}
Lanusse F.,  Ma Q.,  Li N.,  Collett T.~E.,  Li C.-L.,  Ravanbakhsh S.,
  Mandelbaum R.,   Póczos B.,  2017, \mn@doi [Monthly Notices of the Royal
  Astronomical Society] {10.1093/mnras/stx1665}, 473, 3895

\bibitem[\protect\citeauthoryear{Liang et~al.,}{Liang et~al.}{2010}]{Liang2010}
Liang Y.~C.,  et~al., 2010, \mn@doi [Monthly Notices of the Royal Astronomical
  Society] {10.1111/j.1365-2966.2010.16891.x}, 409, 213

\bibitem[\protect\citeauthoryear{Liu, Xia, Mao, Wu  \& Deng}{Liu
  et~al.}{2008}]{LiuFS2008}
Liu F.~S.,  Xia X.~Y.,  Mao S.,  Wu H.,   Deng Z.~G.,  2008, \mn@doi [Monthly
  Notices of the Royal Astronomical Society]
  {10.1111/j.1365-2966.2007.12818.x}, 385, 23

\bibitem[\protect\citeauthoryear{Lowe}{Lowe}{1999}]{1999Object}
Lowe D.~G.,  1999, in Proc of IEEE International Conference on Computer Vision.

\bibitem[\protect\citeauthoryear{{Lupton}, {Gunn}, {Ivezi{\'c}}, {Knapp}  \&
  {Kent}}{{Lupton} et~al.}{2001}]{2001ASPC..238..269L}
{Lupton} R.,  {Gunn} J.~E.,  {Ivezi{\'c}} Z.,  {Knapp} G.~R.,   {Kent} S.,
  2001, in {Harnden} F.~R. J.,  {Primini} F.~A.,   {Payne} H.~E.,  eds,
  Astronomical Society of the Pacific Conference Series Vol. 238, Astronomical
  Data Analysis Software and Systems X. p.~269 (\mn@eprint {arXiv}
  {astro-ph/0101420})

\bibitem[\protect\citeauthoryear{Martin et~al.,}{Martin
  et~al.}{2019}]{MartinKaviraj-137}
Martin G.,  et~al., 2019, Monthly Notices of the Royal Astronomical Society,
  p.~1

\bibitem[\protect\citeauthoryear{Matthews, Driel  \& Ragaigne}{Matthews
  et~al.}{2001}]{MatthewsDriel-149}
Matthews L.~D.,  Driel W.~V.,   Ragaigne D.~M.,  2001, Astronomy and
  Astrophysics, 365

\bibitem[\protect\citeauthoryear{McGaugh}{McGaugh}{1996}]{Mcgaugh-114}
McGaugh S.~S.,  1996, Monthly Notices of the Royal Astronomical Society, 280,
  337

\bibitem[\protect\citeauthoryear{McGaugh, Schombert  \& Bothun}{McGaugh
  et~al.}{1995}]{McgaughSchombert-113}
McGaugh S.,  Schombert J.,   Bothun G.,  1995, arXiv preprint astro-ph/9501085

\bibitem[\protect\citeauthoryear{Monnier~Ragaigne, Van~Driel, Schneider,
  Jarrett  \& Balkowski}{Monnier~Ragaigne
  et~al.}{2003a}]{MonnierRagaigneVanDriel-96}
Monnier~Ragaigne D.,  Van~Driel W.,  Schneider S.~E.,  Jarrett T.~H.,
  Balkowski C.,  2003a, Astronomy \& Astrophysics, 405, 99

\bibitem[\protect\citeauthoryear{Monnier~Ragaigne, van Driel, O'Neil,
  Schneider, Balkowski  \& Jarrett}{Monnier~Ragaigne
  et~al.}{2003b}]{MonnierRagaignevanDriel-59}
Monnier~Ragaigne D.,  van Driel W.,  O'Neil K.,  Schneider S.~E.,  Balkowski
  C.,   Jarrett T.~H.,  2003b, Astronomy \& Astrophysics, 408, 67

\bibitem[\protect\citeauthoryear{Monnier~Ragaigne, van Driel, Schneider,
  Balkowski  \& Jarrett}{Monnier~Ragaigne
  et~al.}{2003c}]{MonnierRagaignevanDriel-60}
Monnier~Ragaigne D.,  van Driel W.,  Schneider S.~E.,  Balkowski C.,   Jarrett
  T.~H.,  2003c, Astronomy and Astrophysics, 408, 465

\bibitem[\protect\citeauthoryear{O~Neil \& Bothun}{O~Neil \&
  Bothun}{2000}]{ONeilBothun-117}
O~Neil K.,  Bothun G.,  2000, The Astrophysical Journal, 529, 811

\bibitem[\protect\citeauthoryear{O'Neil, Bothun  \& Cornell}{O'Neil
  et~al.}{1997}]{O'NeilBothun-106}
O'Neil K.,  Bothun G.~D.,   Cornell M.~E.,  1997, The Astronomical Journal,
  113, 1212

\bibitem[\protect\citeauthoryear{O'Neil, Bothun, Van~Driel  \& Ragaigne}{O'Neil
  et~al.}{2004}]{O'NeilBothun-125}
O'Neil K.,  Bothun G.,  Van~Driel W.,   Ragaigne D.~M.,  2004, Astronomy \&
  Astrophysics, 428, 823

\bibitem[\protect\citeauthoryear{O'Shea \& Nash}{O'Shea \&
  Nash}{2015}]{o2015introduction}
O'Shea K.,  Nash R.,  2015, arXiv preprint arXiv:1511.08458

\bibitem[\protect\citeauthoryear{Pasquet, Bertin, Treyer, Arnouts  \&
  Fouchez}{Pasquet et~al.}{2019}]{Pasquet:2018qlv}
Pasquet J.,  Bertin E.,  Treyer M.,  Arnouts S.,   Fouchez D.,  2019, \mn@doi
  [Astron. Astrophys.] {10.1051/0004-6361/201833617}, 621, A26

\bibitem[\protect\citeauthoryear{Redmon \& Farhadi}{Redmon \&
  Farhadi}{2017}]{RedmonFarhadi-78}
Redmon J.,  Farhadi A.,  2017, YOLO9000: better, faster, stronger

\bibitem[\protect\citeauthoryear{Redmon \& Farhadi}{Redmon \&
  Farhadi}{2018}]{RedmonFarhadi-79}
Redmon J.,  Farhadi A.,  2018, arXiv preprint arXiv:1804.02767

\bibitem[\protect\citeauthoryear{Redmon, Divvala, Girshick  \& Farhadi}{Redmon
  et~al.}{2016}]{RedmonDivvala-140}
Redmon J.,  Divvala S.,  Girshick R.,   Farhadi A.,  2016, You Only Look Once:
  Unified, Real-Time Object Detection

\bibitem[\protect\citeauthoryear{Ren, He, Girshick  \& Sun}{Ren
  et~al.}{2015}]{RenHe-65}
Ren S.,  He K.,  Girshick R.,   Sun J.,  2015, arXiv preprint arXiv:1506.01497

\bibitem[\protect\citeauthoryear{Robitaille~Thomas et~al.,}{Robitaille~Thomas
  et~al.}{2013}]{robitaille2013streicher}
Robitaille~Thomas P.,  et~al., 2013, Astropy: A community Python package for
  astronomy//Astronomy \& Astrophysics. X, 558, A33

\bibitem[\protect\citeauthoryear{Trachternach, Bomans, Haberzettl  \&
  Dettmar}{Trachternach et~al.}{2006}]{TrachternachBomans-120}
Trachternach C.,  Bomans D.~J.,  Haberzettl L.,   Dettmar R.-J.,  2006,
  Astronomy \& Astrophysics, 458, 341

\bibitem[\protect\citeauthoryear{Van Der~Walt, Colbert  \& Varoquaux}{Van
  Der~Walt et~al.}{2011}]{van2011numpy}
Van Der~Walt S.,  Colbert S.~C.,   Varoquaux G.,  2011, Computing in science \&
  engineering, 13, 22

\bibitem[\protect\citeauthoryear{Van~Rossum \& Drake}{Van~Rossum \&
  Drake}{2009}]{van2009python}
Van~Rossum G.,  Drake F.,  2009, Scotts Valley, CA

\bibitem[\protect\citeauthoryear{Virtanen et~al.,}{Virtanen
  et~al.}{2020}]{virtanen2020scipy}
Virtanen P.,  et~al., 2020, Nature methods, 17, 261

\bibitem[\protect\citeauthoryear{Walmsley et~al.,}{Walmsley
  et~al.}{2019}]{10.1093/mnras/stz2816}
Walmsley M.,  et~al., 2019, \mn@doi [Monthly Notices of the Royal Astronomical
  Society] {10.1093/mnras/stz2816}, 491, 1554

\bibitem[\protect\citeauthoryear{Wes, van~der Walt  \& Millman}{Wes
  et~al.}{2010}]{wes2010proc}
Wes M.,  van~der Walt S.,   Millman J.,  2010, Proc. 9th Python in Sci. Conf

\bibitem[\protect\citeauthoryear{York et~al.,}{York
  et~al.}{2000}]{YorkAdelman-147}
York D.~G.,  et~al., 2000, The Astronomical Journal, 120, 1579

\bibitem[\protect\citeauthoryear{Zhong, Liang, Liu, Hammer, Hu, Chen, Deng  \&
  Zhang}{Zhong et~al.}{2008}]{ZhongLiang-105}
Zhong G.~H.,  Liang Y.~C.,  Liu F.~S.,  Hammer F.,  Hu J.~Y.,  Chen X.~Y.,
  Deng L.~C.,   Zhang B.,  2008, Monthly Notices of the Royal Astronomical
  Society, 391, 986

\bibitem[\protect\citeauthoryear{and A.~M. Price-Whelan et~al.,}{and A.~M.
  Price-Whelan et~al.}{2018}]{2018}
and A.~M. Price-Whelan et~al., 2018, \mn@doi [The Astronomical Journal]
  {10.3847/1538-3881/aabc4f}, 156, 123

\bibitem[\protect\citeauthoryear{de Blok, McGaugh  \& van~der Hulst}{de~Blok
  et~al.}{1996}]{deBlokMcgaugh-150}
de Blok W. J.~G.,  McGaugh S.~S.,   van~der Hulst J.~M.,  1996, \mn@doi
  [Monthly Notices of the Royal Astronomical Society] {10.1093/mnras/283.1.18},
  283, 18

\makeatother
\end{thebibliography}
	
	%%%%%%%%%%%%%%%%%%%%%%%%%%%%%%%%%%%%%%%%%%%%%%%%%%
%\bsp	% typesetting comment
\label{lastpage}
\end{document}